\documentclass[10pt,conference]{IEEEtran}

\usepackage{amssymb}
\usepackage{cite}
\usepackage[dvips]{graphicx}
\usepackage[cmex10]{amsmath}
\usepackage{array}

\def\argmax{\mathop{\rm arg\,max}}

\begin{document}

\title{Optimal Detector for Channels with Non-Gaussian Interference}

\author{\IEEEauthorblockN{Jungwon Lee\IEEEauthorrefmark{1},
Dimitris Toumpakaris\IEEEauthorrefmark{2}, and 
Hui-Ling Lou\IEEEauthorrefmark{1}}
\IEEEauthorblockA{\IEEEauthorrefmark{1}Marvell Semiconductor Inc., 5488 Marvell Ln, Santa Clara, CA 95054}
\IEEEauthorblockA{\IEEEauthorrefmark{2}Wireless Telecommunications Laboratory, Department of Electrical and Computer Engineering \\ University of Patras, Rio, Greece 265 00\\
Emails: jungwon@stanfordalumni.org, dtouba@upatras.gr, lou@stanfordalumni.org}}

\maketitle

\begin{abstract}
The detection problem in the Gaussian interference channel is addressed, when transmitters employ non-Gaussian schemes designed for the single-user Gaussian channel. 
A structure consisting of a separate symbol-by-symbol detector and a hard decoder is considered.
Given this structure, an optimal detector is presented that is compared to an interference-unaware conventional detector, an interference-aware successive interference cancellation (SIC) detector, and a minimum-distance detector. It is demonstrated analytically and by simulation that the optimal detector outperforms both the conventional and the SIC detector, and that it attains decreasing symbol error rates even in the presence of strong interference. Moreover, the minimum-distance detector performs almost as well as the optimal detector in most scenarios and is significantly less complex.
\end{abstract}

\section{Introduction}
The Gaussian interference channel has significant practical interest, because it can model real-world communication systems. Examples include cellular systems where neither base stations nor mobile stations coordinate with each other and bundles of digital subscriber lines (DSL) without any real-time coordination among modems of different users.
The interference channel is one of the most challenging channels studied in Information Theory, and the complete characterization of its capacity region remains an open problem \cite{Carleial75, HK81, Sason04, RH04, ETW07, CJS09}. 

A significant part of the research activity on the interference channel has focused on determining its capacity and the coding schemes with which it can be achieved. Another area of considerable activity is power allocation for frequency-selective Gaussian interference channels
\cite{YGC02, CYMVB06, LSC02}.
For power allocation it is typically assumed that each user employs encoding and decoding schemes designed for the single-user Gaussian channel. Moreover, the interference is treated as independent identically distributed (i.i.d.) Gaussian, similar to background noise. Therefore, it is assumed that the transmitters and the receivers are the same as for the single-user Gaussian channel. 

Thus, a 
gap seems to exist between research on the capacity region and research on power allocation for Gaussian interference channels. Given the current technology, near-capacity-achieving encoding and decoding schemes are very complex to implement in practical communication systems \cite{HK81, ETW07}. On the other hand, power allocation is already performed in some systems. It is therefore of interest to improve the performance of the latter, while keeping complexity at a reasonable level.

In this paper, an attempt is made to bridge this gap. 
It is assumed that simple encoding schemes designed for single-user Gaussian channels are employed. On the other hand, decoding schemes are explored that do not necessarily treat interference as i.i.d. Gaussian. 

Soft or hard estimates can be used for decoding. 
Soft decoding performs better than hard decoding. However, hard decoding is simpler because, unlike the decoding algorithm, the detection does not depend on the employed encoding scheme. Moreover, hard decoding can provide some insights on the decoding of single-user-coded signals in the presence of interference. In order to perform the optimization of the detector in a general context, without assuming a specific encoding scheme, hard decoding is assumed in this paper. Soft decoding is considered in \cite{Lee08_1}.

The focus of this paper is the development of an interference-aware detector for hard decoding receivers that can be applied to any coding scheme designed for single-user Gaussian channels. A conventional detector is first examined briefly that does not take the statistics of interference into account. 
Then an interference-aware successive interference cancellation (SIC) detector is described, followed by a minimum-distance detector. Finally, the optimal detector is derived. The detectors are compared in terms of the symbol error rate (SER) analytically and by simulation. It is shown that the optimal detector outperforms significantly both the conventional and the SIC detector. Moreover, the performance of the minimum-distance detector, that is considerably less complex than the optimal one, is optimal for most values of the signal-to-noise and signal-to-interference ratio. 

The remainder of this paper is organized as follows. The model of the system that is considered is presented in Section \ref{sec:system_model} followed by the four detectors in Section \ref{sec:Detector}. The performance of the detectors is compared analytically in Section \ref{sec:comparison} and using simulations in Section \ref{sec:simulation_results}. 

\section{System Model}
\label{sec:system_model}
\begin{figure*}[!t]
\centering
\includegraphics[width=7in]{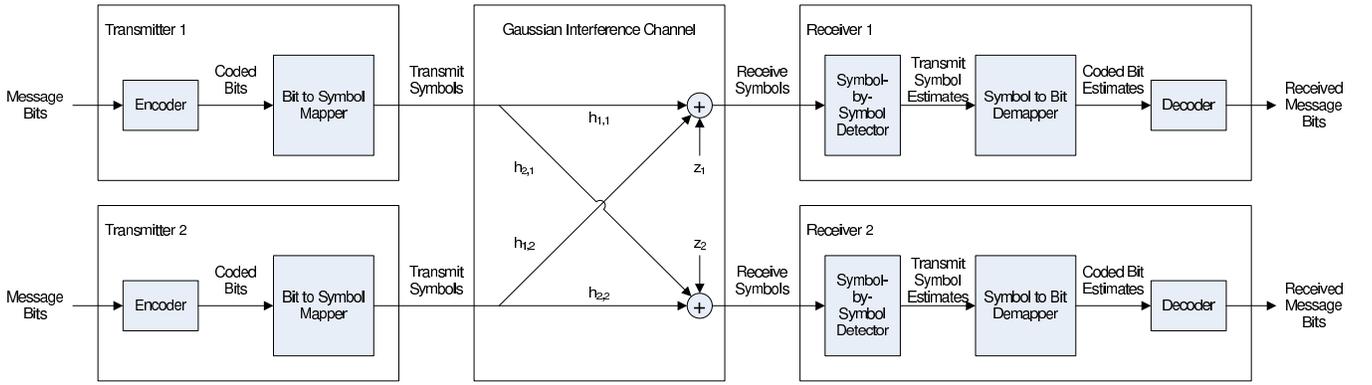}
\caption{System Model}
\label{fig:system_model}
\end{figure*}

A two-user Gaussian interference channel is considered
\begin{eqnarray}
y_1[m] & = & h_{1,1}[m] x_1[m] +h_{1,2}[m] x_2[m] + z_1[m], \nonumber \\
y_2[m] & = & h_{2,1}[m] x_1[m] +h_{2,2}[m] x_2[m] + z_2[m],
\label{eq:receive_signal_model}
\end{eqnarray}
where $x_j[m]$ is the signal of transmitter $j$ at time $m$, $h_{i,j}[m]$ is the gain of the channel from transmitter $j$ to receiver $i$, and $z_i[m]$ is the background noise of receiver $i$, for $i=1,2$ and $j=1,2$. 
The background noise is modeled as a real Gaussian random variable with variance $\tilde{\sigma}_z^2$ for baseband systems and circularly symmetric complex Gaussian with variance $2\tilde{\sigma}_z^2$ for passband systems.
It is assumed that each receiver $i$ knows all channel gains $h_{i,j}[m]$,
which is a typical assumption for any advanced receiver for interference channels \cite{ETW07}.
The model
is shown in Fig.~\ref{fig:system_model}. %
As can be seen, a practical transmitter is considered where each symbol is a point of a finite signal constellation. Examples include pulse-amplitude modulation (PAM), quadrature-amplitude modulation (QAM), and phase shift keying (PSK). 
In the following, the time index $m$ will be omitted for simplicity whenever there is no potential for confusion.

\section{Detectors for the interference channel for single-user encoding schemes}
\label{sec:Detector}


For the receiver, a suboptimal, but practically viable and frequently employed approach, is to design a symbol-by-symbol detector and a decoder separately as in Fig.~\ref{fig:system_model}. This structure is used in this paper and allows optimization of the detector independently from the decoder. 

\subsection{Conventional Detector}

In the absence of interference, the optimal detector for $x_1$ given a hard-decoding receiver 
maps the received symbol to the closest point of the constellation of transmitter 1 scaled by $h_{1,1}$.
When interference is present, it is treated by the detector as part of the background noise. Specifically, in (\ref{eq:receive_signal_model}), the interference term $h_{1,2} x_2$ is lumped together with the background noise term $z_1$. 
The decision regions of the conventional detector for 2-PAM are shown in Fig.~\ref{fig:decision_region}.

In many cases, interference can be safely assumed to be Gaussian. For example, in DSL, combined interference from many users can be considered Gaussian. However, in other cases, the modeling of interference as a Gaussian signal may not be accurate. For example, in the case of a cellular system employing fractional frequency reuse, the number of non-negligible interferers can be restricted to a small value. Even when the interference comes from multiple users as in DSL, it can improve the system performance to treat a few strong interferers separately from the rest, which can still be approximated as Gaussian. Thus, in the following, detectors are considered that exploit the statistics of the interference, which can be different from Gaussian.

\subsection{Interference-Aware Detector}

The conventional detector is optimal if the unwanted signals (\emph{i.e.,} the sum of noise and interference) are Gaussian. 
When the interfering users employ finite constellations, the distribution of the unwanted signal is not Gaussian. The performance of the receiver can be improved by using a detector that takes into account the actual distribution of the interference.

It is assumed that the modulation scheme of transmitter 2 is known to receiver 1 and that the gain $h_{1,2}$ is known during each time instant $m$. Therefore, the distribution of the received interference signal $h_{1,2} x_2$ is also known.
 

\subsubsection{SIC Detector}
For the development of an interference-aware detector, the concept of successive interference cancellation (SIC) can be borrowed from the multiple access channel (MAC). 
Even though receiver 1 is interested in $x_1$, it can obtain an estimate $\hat{x}_2$ of $x_2$ first, by treating $x_1$ as Gaussian. Then the receiver can detect $x_1$ from $y_1 - h_{1,2}\hat{x}_2$. The decision regions for the SIC detector when 2-PAM is used by both transmitters are shown in Fig.~\ref{fig:decision_region}.

The performance of the SIC detector will be good when the power of the interference is much larger than the power of the desired signal (small SIR). On the other hand, for large SIR, the conventional detector may perform better than the SIC detector. Based on this observation, an \emph{ordered} SIC detector can also be considered. For example, when 2-PAM is used by both transmitters, $x_1$ can be directly estimated for SIR $\geq$ 0 dB, whereas the SIC decoder can be used for SIR $<$ 0 dB. In the description of the minimum-distance detector that follows, it is shown that this ordered SIC can attain the same performance as the minimum-distance detector for the case of 2-PAM. However, in general, for other constellation sizes, the optimal ordering sequence is not obvious.

\subsubsection{Minimum-Distance Detector}
The SIC detector may be affected adversely by error propagation.
Because symbol-by-symbol detection does not benefit from coding, this can lead to severe performance degradation at small interference-to-noise ratios (INR). To avoid error propagation, $x_1$ and $x_2$ can be estimated jointly rather than sequentially. Detection of $x_1$ can be accomplished in three steps. First, a combined received constellation point $h_{1,1} x_1 + h_{1,2} x_2$ is formed for each pair $(x_1,x_2) \in S_1 \times S_2$ where $S_i$ are the transmit signal constellations. 
The detector then determines the combined constellation point 
that is closest to $y_1$ among all the points in $h_{1,1} S_1 + h_{1,2} S_2$ and the corresponding $(\hat{x}_1, \hat{x}_2)$. Finally, $\hat{x}_1$ from $(\hat{x}_1, \hat{x}_2)$ is chosen as the estimate of $x_1$. It can be easily shown that this decision rule minimizes the probability of mis-detection of the pair $(\hat{x}_1, \hat{x}_2)$ when all constellation points in the finite alphabet sets $S_1$ and $S_2$ are equiprobable.

As in the SIC detector, $x_2$ is also estimated, but only through the joint estimation of $x_1$ and $x_2$. However, unlike the SIC detector, the estimate of $x_1$ does not rely directly on the estimate of $x_2$. This eliminates the effect of error propagation in low INR.
Therefore, $\hat{x}_1$ may be correct even if $\hat{x}_2$ is incorrect. In summary, the minimum-distance detector estimates both $x_1$ and $x_2$, but only uses $\hat{x}_1$; the conventional detector only estimates $x_1$; and the SIC detector estimates $x_2$ and $x_1$ sequentially.



\subsubsection{Optimal Maximum-Likelihood (ML) Detector}

The minimum-distance detector is not optimal with respect to minimizing the probability of mis-detection of the desired signal. It can be shown that the optimal, maximum-likelihood (ML) detector for $x_1$ given the received signal $y$ is given by 
\begin{equation}
\hat{x}_1(y_1) = \argmax_{x_1} \sum_{x_2} \exp ( -|y_1 - h_{1,1} x_1 - h_{1,2} x_{2}|^2/2 \bar{\sigma}_Z^2),
\label{eq:opt_ML}
\end{equation}
where the summation occurs over all the symbols of the constellation of the interferer.
Compared to the minimum-distance detector, the optimal ML detector is rather complex in that it requires the calculation of the sum of exponential functions and the calculation of the Euclidean distance from the received signal to all combined signal constellation points. As will be seen in the following, except for channels where the power of the interference is comparable to the signal power and the SNR is small, the performance of the minimum-distance detector is almost as good as that of the optimal detector. Therefore, the significantly less complex minimum-distance detector can be employed instead.

\begin{figure}[!t]
\centering
\includegraphics[width=3.3in]{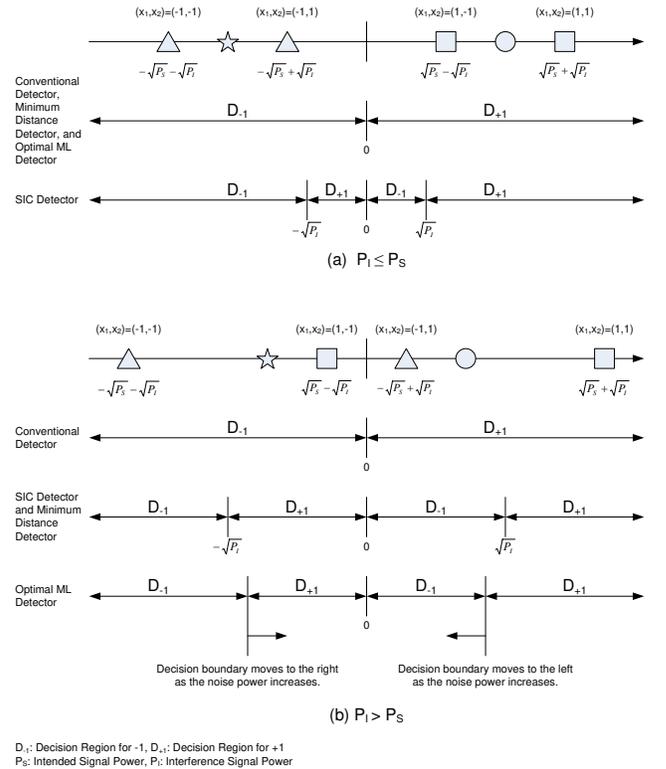}
\caption{Decision regions of the decoders for weak and strong interference.}
\label{fig:decision_region}
\end{figure}

The decision regions of the four decoders are shown in Fig.~\ref{fig:decision_region}, for the case when both transmitters use 2-PAM.
As can be seen, when $\mbox{SIR}<0$ dB, the decision regions of the optimal ML detector change slightly compared to the minimum-distance detector. This happens because, unlike the minimum-distance detector, more than one combined symbols contribute to each decision as can be seen in (\ref{eq:opt_ML}). The difference is more pronounced for small SNR and disappears as the noise variance decreases and the exponential terms decay faster. It can also be seen that, for 2-PAM, the decision regions of the ordered SIC decoder will be the same as for the minimum-distance decoder. However, this does not hold for arbitrary PAM constellation size.

The first three detectors are similar to detectors for a multi-input-single-output (MISO) system with 2 transmit antennas and 1 receive antenna. Specifically, the conventional detector, the SIC detector, and the minimum-distance detector correspond to a linear equalizer, a decision feedback equalizer, and a maximum likelihood detector for a MISO system using more than one spatial streams. Nevertheless, in practice, only one spatial stream is transmitted in a MISO system, using either space-time coding or beam-forming because use of multiple streams does not lead in capacity gains, and requires more complex receivers compared to the single-stream case. 


\section{Comparison of the detectors}
\label{sec:comparison}


\subsection{Symbol Error Rate}
When 2-PAM is used by both transmitters, the SER of the first three detectors can be derived easily using Fig.~\ref{fig:decision_region}. 
However, the resulting expression is quite long and complicated. Thus, only approximate SER expressions, based on the nearest neighbor union bound (NNUB) approach \cite{Cioffi379A}, are given below. The derivation of the expressions is based on finding the minimum distance between points of the constellation $h_{1,1} S_1 + h_{1,2} S_2$ that decode to different values of $x_1$, and is omitted here for brevity. The exact SER expressions and the derivations can be found in \cite{LTL09}. The SER expressions for the optimal ML detector are even more complicated, even for the 2-PAM case. Therefore, in this paper, the expressions of the minimum-distance detector are used as an upper bound for the performance of the optimal detector.

The SER for the conventional detector when 2-PAM is used by both transmitters is approximately equal to
\begin{equation}
P_{e,\mbox{conv}} \approx \left\{ \begin{array}{ll}
\frac{1}{2} Q \left( \sqrt{\mbox{SNR}} - \sqrt{\mbox{INR}} \right), & \mbox{SIR} \geq 1 \\
\frac{1}{2} - \frac{1}{2} Q \left( \sqrt{\mbox{INR}} - \sqrt{\mbox{SNR}} \right), & SIR < 1,
\end{array} \right.
\label{eq:SER_conv}
\end{equation}
where $Q(x) = \frac{1}{\sqrt{2\pi}} \int_{x}^{\infty} e^{-t^2/2} dt$ is the Q-function.
The SER for the SIC detector is
\begin{equation}
P_{e,\mbox{SIC}} \approx \left\{ \begin{array}{ll}
\frac{1}{2} Q \left( \sqrt{\mbox{SNR}} - 2 \sqrt{\mbox{INR}} \right), & \mbox{SIR} \geq 4 \\
\frac{1}{2} - \frac{1}{2} Q \left( \sqrt{\mbox{SNR}} - \sqrt{\mbox{INR}} \right), & \frac{9}{4} \leq \mbox{SIR} < 4 \\
\frac{1}{2} - \frac{1}{2} Q \left( 2\sqrt{\mbox{INR}} - \sqrt{\mbox{SNR}} \right), & 1 \leq \mbox{SIR} < \frac{9}{4} \\
\frac{1}{2} Q \left( \sqrt{\mbox{INR}} - \sqrt{\mbox{SNR}} \right), & \frac{1}{4} \leq \mbox{SIR} < 1 \\
Q \left( \sqrt{\mbox{SNR}} \right), & \mbox{SIR} < \frac{1}{4}. \\
\end{array} \right. 
\label{eq:SER_SIC}
\end{equation}
Lastly, the SER of the minimum-distance detector is given by
\begin{equation}
P_{e,\mbox{MD}} \approx \left\{ \begin{array}{ll}
\frac{1}{2} Q \left( \sqrt{\mbox{SNR}} - \sqrt{\mbox{INR}} \right), & \mbox{SIR} \geq 1 \\
\frac{1}{2} Q \left( \sqrt{\mbox{INR}} - \sqrt{\mbox{SNR}} \right), & \frac{1}{4} \leq \mbox{SIR} < 1 \\
Q \left( \sqrt{\mbox{SNR}} \right), & \mbox{SIR} < \frac{1}{4}. \\
\end{array} \right. 
\label{eq:SER_opt}
\end{equation}
From (\ref{eq:SER_opt}) it can be seen that the minimum-distance detector behaves roughly as follows. For SIR $\geq 1$, the minimum distance of the constellation of the desired signal is reduced because of interference. For $\frac{1}{4} \leq$ SIR $< 1$, the detector behaves as if the minimum distance in the constellation of the interference were reduced by the desired signal. For SIR $< \frac{1}{4}$, the interference is very strong and its effect on the SER is negligible. This agrees with the information theoretic results for the strong interference channel.
Comparing (\ref{eq:SER_conv}) and (\ref{eq:SER_opt}), it can be deduced that the minimum-distance detector outperforms the conventional detector for SIR $<$ 1. In fact, the SER of the conventional detector exceeds $1/4$ for SIR $<$ 1. From (\ref{eq:SER_SIC}) and (\ref{eq:SER_opt}), for SIR $\geq$ 1, the SER of the minimum-distance detector is smaller than the SER of the SIC detector. The performance of the SIC detector is particularly bad in the region $1 \leq$ SIR $< 4$.


Although only 2-PAM was considered, the SER for other modulation schemes can be derived in a similar way using the NNUB approach. However, the resulting expressions are typically quite complicated. It can be shown that the performance of the conventional detector is the same as that of the minimum-distance detector when the SIR exceeds a threshold value SIR$_{th}$. On the contrary, the conventional detector exhibits an error floor for SIR $<$ SIR$_{th}$. For 2-PAM, SIR$_{th} = 1$. The following proposition holds for any PAM constellation size. The proof is given in \cite{LTL09}. 

\emph{Proposition} 1. For the 2-user interference channel where both transmitters employ PAM, the SER of the conventional detector reaches an error floor when the SIR is below a threshold SIR$_{th}$. Mathematically,
for $\mbox{SIR} > \mbox{SIR}_{th}$, 
$\mbox{P}_{e,\mbox{opt}} = \mbox{P}_{e,\mbox{conv}} \rightarrow 0 \mbox{ as SNR} \rightarrow \infty.$
For $\mbox{SIR} \leq \mbox{SIR}_{th}$, 
$\mbox{P}_{e,\mbox{opt}} \rightarrow 0 \mbox{ as SNR} \rightarrow \infty,$
except for a finite number of SIR values.
On the other hand,
$\mbox{P}_{e,\mbox{conv}} \rightarrow \mbox{P}_{e,\mbox{floor}} \neq 0 \mbox{ as SNR} \rightarrow \infty,$
\emph{i.e.,} the SER reaches a non-zero error floor.
\begin{equation}
\mbox{SIR}_{th} = (M^2_1-1)(M_2-1)/(M_2+1),
\label{eq:threshold}
\end{equation}
where $M_1$ and $M_2$ is the constellation size of transmitter 1 (desired) and transmitter 2 (interfering), respectively.

Similarly, the SER of the SIC detector reaches an error floor when the SIR is between two threshold values SIR$_{1}$ and SIR$_{2}$. 
$\mbox{P}_{e,\mbox{SIC}} \rightarrow \mbox{P}_{e,\mbox{floor,SIC}} \neq 0 \mbox{ as SNR} \rightarrow \infty,$
when SIR$_1 \leq \mbox{SIR} \leq $ SIR$_2$.
\begin{eqnarray}
\mbox{SIR}_1 & = & (M_1+1)/\left\{(M_1-1)(M^2_2-1)\right\}, \ \mbox{and} \nonumber \\
\mbox{SIR}_2 & = & 4(M^2_1-1)(M_2-1)/(M_2+1) = 4 \mbox{SIR}_{th}.
\end{eqnarray}


Although only PAM was considered for the analysis, similar conclusions hold for passband systems with QAM.

\subsection{Complexity}

At most $M_1-1$ comparisons are required by the conventional detector, compared to at most $M_1 + M_2 - 2$ for
the SIC detector. The maximum number of comparisons for the minimum-distance detector is equal to $M_1 M_2 - 1$. This increase in complexity is not important when compared to the complexity of other parts of the receiver (such as a Viterbi decoder that may follow the detector). Moreover, in practical implementations, the number of comparisons is reduced substantially by optimizing the design of the slicer. Finally, the optimal ML detector is significantly more complex. Nevertheless, as is shown in the following, it can be replaced by the minimum-distance detector in most cases.

\section{Simulation Results}
\label{sec:simulation_results}

\begin{figure}[!t]
\centering
\includegraphics[width=3.2in]{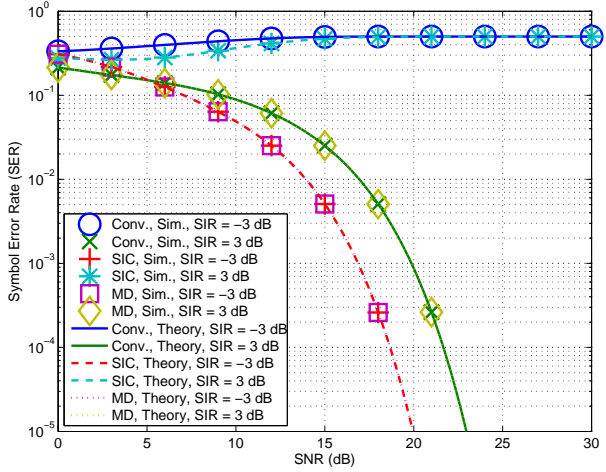}
\caption{Symbol error rate (SER) for the conventional detector, the SIC detector, and the minimum-distance detector for a two-user Gaussian interference channel when 2-PAM is used by both transmitters.}
\label{fig:SER_PAM_2_2}
\end{figure}

The performance of the 
detectors is evaluated by Monte Carlo simulation. 
Fig.~\ref{fig:SER_PAM_2_2} shows the SER when 2-PAM is used by both transmitters. It can be seen that the SER of the conventional detector reaches an error floor when SIR=$-$3 dB. However, the optimal detector performs very well even at that very strong interference environment. In fact, as can be seen in the figure, the minimum-distance detector performs better for SIR = $-3$ dB than for SIR=3 dB for a given SNR. The SIC detector is as good as the optimal detector for SIR=$-$3 dB, but its performance diverges when SIR=3 dB. As can also be seen in Fig.~\ref{fig:SER_PAM_2_2}, the expressions derived using the NNUB approximation are very close to the simulation results.

\begin{figure}[!t]
\centering
\includegraphics[width=3.2in]{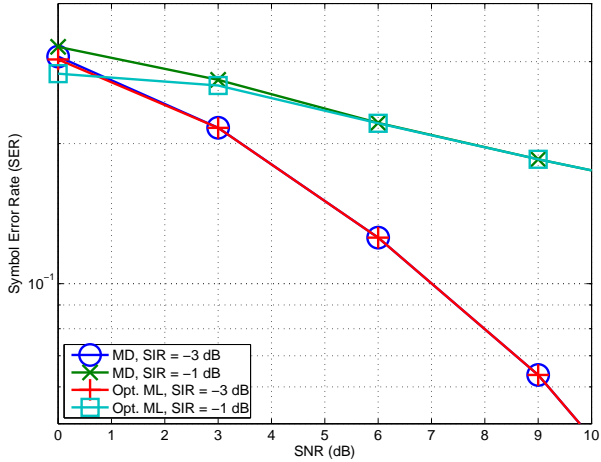}
\caption{Symbol error rate (SER) for the minimum-distance detector and the optimal ML detector for a two-user Gaussian interference channel when 2-PAM is used by both transmitters.}
\label{fig:SER_PAM_2_2_MD_opt}
\end{figure}

In Fig.~\ref{fig:SER_PAM_2_2_MD_opt} the SER for the minimum-distance detector and the optimal ML detector is compared. The minimum-distance detector performs almost as well as the optimal ML detector except for very low SNRs of less than 3 dB when SIR = $-1$ dB, \emph{i.e.}, the power of the interference is comparable to that of the desired signal. 

\begin{figure}[!t]
\centering
\includegraphics[width=3.2in]{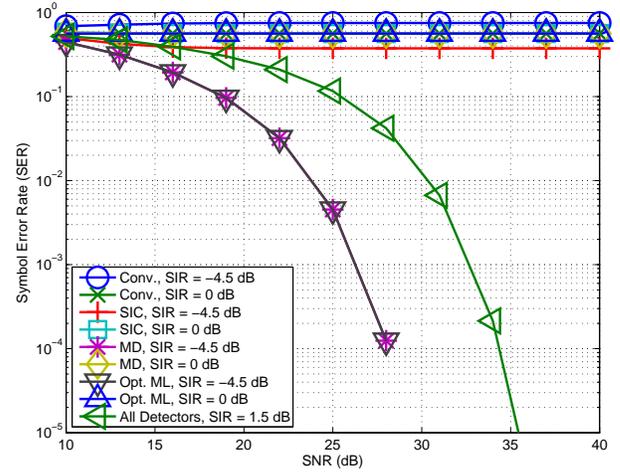}
\caption{Symbol error rate (SER) for all 4 detectors when 4-PAM is used by both transmitters.}
\label{fig:SER_PAM_4_4}
\end{figure}

The SER for the case where 4-PAM is used by both transmitters is plotted in Fig.~\ref{fig:SER_PAM_4_4}. It can be seen that the minimum-distance detector works well for both SIR=3 dB and SIR=$-3$ dB, whereas the conventional detector and the SIC detector fail. This agrees with Proposition 1, since, from (\ref{eq:threshold}), SIR$_{th}\approx$ 9.54 dB and $-9.54 \ \mbox{dB}=\mbox{SIR}_1<\mbox{SIR}<\mbox{SIR}_2 = 14.3$ dB.. Therefore, use of the minimum-distance detector enables employing 4-PAM over the interference channel that would otherwise be impossible. 


\begin{figure}[!t]
\centering
\includegraphics[width=3.2in]{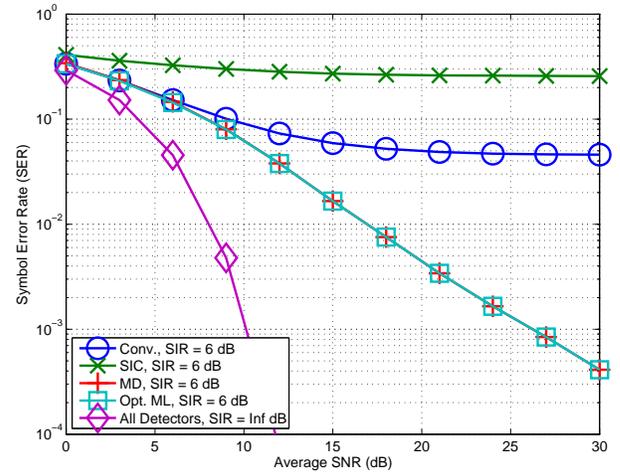}
\caption{SER for all 4 detectors in a Rayleigh fading channel when 4-QAM is used by both transmitters.}
\label{fig:SER_QAM}
\end{figure}

Finally, in Fig.~\ref{fig:SER_QAM}, the SER is depicted when 4-QAM is used by both transmitters and transmission occurs over a fading channel. The distribution of each channel gain is assumed to be i.i.d. complex Gaussian. 
The conventional and the SIC detector suffer significantly from fading even for an average SIR of 6 dB. On the other hand, the minimum-distance detector does not exhibit an error floor even when the average SIR is equal to $-6$ dB. 

\end{document}